\documentclass[twocolumn,preprintnumbers,amsmath,amssymb,superscriptaddress, altaffilletter, subeqn,prl, maxbibnames=99]{revtex4-2}
\usepackage{physics}
\usepackage{dcolumn}
\usepackage{bm}					
\usepackage{graphicx}
\usepackage{amsmath}
\usepackage[dvipsnames]{xcolor}
\usepackage{float}
\usepackage{comment}

\begin{document}

\title{Ab-initio density-matrix approach to exciton coherence:\\ phonon scattering, Coulomb interactions and radiative recombination}

\author{Tomer Amit}

\affiliation{Department of Molecular Chemistry and Materials Science, Weizmann Institute of Science, Rehovot 7610001, Israel}

\author{Guy Vosco}

\affiliation{Department of Molecular Chemistry and Materials Science, Weizmann Institute of Science, Rehovot 7610001, Israel}

\author{Mauro Del Ben}
\affiliation{Applied Mathematics \& Computational Research Division, Lawrence Berkeley National Laboratory, Berkeley, California 94720, United States}

\author{Sivan Refaely-Abramson}
 
\affiliation{Department of Molecular Chemistry and Materials Science, Weizmann Institute of Science, Rehovot 7610001, Israel}

\email[Corresponding author:]{sivan.refaely-abramson@weizmann.ac.il}

\begin{abstract}
Relaxation processes following light excitation in semiconductors are key in materials-based quantum technology applications. These processes are broadly studied in atomically thin transition metal dichalcogenides (TMDs), quasi-two-dimensional excitonic semiconductors in which atomistic design allows for tunable excited-state properties, such as relaxation lifetimes and photo-induced coherence. In this work, we present a density-matrix-based approach to compute exciton relaxation within a many-body ab initio perspective. We expand our previously developed Lindblad density-matrix formalism to capture multi-channel electron-hole pair relaxation processes, including phonon and Coulomb scattering as well as radiative recombination, and study their effect on the time-resolved excited-state propagation. Using monolayer MoSe$_2$ as a prototypical example, we examine many-body effects on the time-dependent dynamics of photoactive excitations, exploring how the electron-hole pair interactions are reflected in variations of the excitation energy, spectral signature, and state coherence. Our method supplies a detailed understanding of exciton relaxation mechanisms in realistic materials, offering a previously unexplored pathway to study excited-state dynamics in semiconductors from first principles.
\end{abstract} 
 
\maketitle
\noindent\textbf{Introduction}

Light-matter interactions in layered semiconducting materials generate excited particles that can relax into long-lived and coherent quantum states~\cite{CohenLouie}. These are of great interest for materials-based quantum information science~\cite{QuantumComputing, kennes2021moire}, with the promise of well-defined quantum coherences upon structural design~\cite{fogler2014high,katzer2023exciton}.
A key ingredient in these relaxation processes are \textit{excitons}, correlated electron-hole pairs bound together by Coulomb interactions, which often serve as the main carriers. Transition metal dichalcogenides (TMDs)~\cite{qiu2013optical, berkelbach2018} are important examples of structurally designable excitonic systems, in which low-energy, strongly-bound, and long-lived excitons are populated following a photoexcitation.  
The efficiency of such population is determined by the time-resolved scattering dynamics of the optical excitation, and advances in the ability to carefully detect these processes allow their study as a function of the underlying material structure~\cite{Ginsberg2020}. Interface composition, interalyer twisting, and the introduction of defects are broadly explored examples of such structural modifications that enhance non-radiative pathways to populate long-lived coherent states through charge separation and spatial localization~\cite{Jin2018, yuan2020twist, mitterreiter2021role, hotger2023spin, troue2023extended, refaely2018defect, amit2022tunable, kundu2023exciton}.

A comprehensive understanding of the relation between exciton relaxation dynamics and structural design demands an accurate theory together with a robust implementation capable of performing numerical simulations for realistic systems beyond toy models. In particular, first-principles computations are essential to provide a detailed understanding of how structural complexity influences exciton transport dynamics~\cite{Qiu2021}.
A common approach to accurately describe structure-specific exciton properties is by the solution of a many-body Bethe Salpeter equation within many-body perturbation theory in the GW-BSE approach~\cite{Hybertsen1986, Rohlfing1998}. Within this framework, Coulomb and exchange interactions between multiple electron and hole states, with varying crystal momentum and spin properties, are computed explicitly. This allows for an accurate consideration of dielectric effects, spin-orbit coupling, and exciton dispersion~\cite{Qiu2015, Cudazzo2015}. Due to rapid progress of algorithms and computational capabilities, many-body perturbation theory became the state-of-the-art approach for calculating excited-state properties in structurally-complex systems~\cite{louie2021discovering}.  
Advances in the field further include the development of GW-BSE-based non-equilibrium and perturbative approaches~\cite{Attaccalite2011, perfetto2022real, perfetto2023real}, allowing for accurate simulations of processes such as transient absorption, time-resolved and angle-resolved photoemission (TR-ARPES), two-photon absorption spectra, and nonlinear spectroscopy, by accounting for temporal evolution of the dielectric function and its consequences in time-resolved photexcitation processes~\cite{perfetto2015nonequilibrium, sangalli2018ab, chan2021giant, sangalli2021excitons}. Recent theory developments further allow for predictive calculations of exciton-exciton scattering through phonons~\cite{antonius2022theory, chen2022first, chan2023exciton, cohen2024phonon}, typically using density functional perturbation theory (DFPT)~\cite{giustino2017electron} within the \textit{ab initio} perspective. Still, taking the wealth of participating particles in the relaxation processes from a GW-BSE perspective is a very challenging task.

\begin{figure}[H]\includegraphics[width=1.0\linewidth]{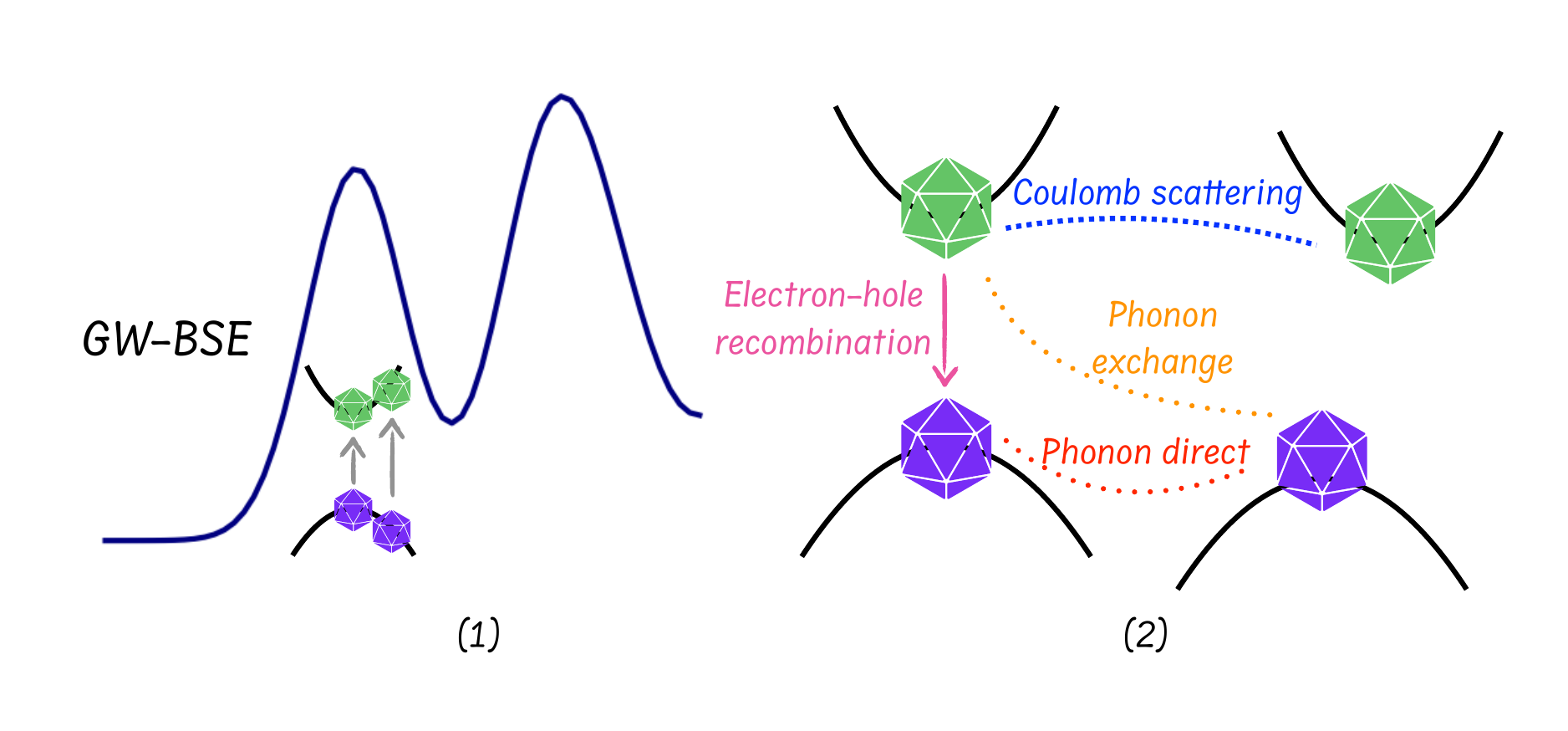}
\caption{Schematic representation of exciton dynamics as described in this paper. (1) Electrons (green) are excited to unoccupied bands, leaving behind holes in an occupied band (purple). GW-BSE computations give rise to an absorption spectrum composed of many-body excitonic states (dark blue). (2) The examined scattering channels. Direct phonon interactions are shown in red, phonon-exchange interactions in orange and Coulomb scattering in blue. Presented interactions correspond to the diagonal case of the studied equation of motion. Electron-hole recombination is shown in pink.}\label{dyn_scheme}
\end{figure}

Recently, we have presented a GW-BSE-based two-particle Linbdlad density-matrix approach~\cite{amit2023ultrafast} to compute direct phonon-induced exciton decomposition due to changes in the time-resolved occupation of the electron-hole pairs composing the many-body excitations. We demonstrated our approach for the case of intralayer and interlayer excitons in TMD heterostructures, showing how intralayer excitons rapidly delocalize across the interface, reducing the spin number of the optically-induced excitations and suggesting a pathway for ultrafast exciton coherence. Here, we explicitly compute the exciton coherence by expanding our approach to include Coulomb interactions between the electron-hole pairs in the density matrix propagation, as well as direct and exchange phonon scattering and radiative electron-hole recombination, as schematically shown in Fig.~\ref{dyn_scheme}. We compute changes in the optically-excited states at ultrafast time scales to study many-body effects on these exciton relaxation channels. Our method and the results obtained with it provide a new first-principles perspective to calculate ultrafast transitions and modifications in exciton characteristics upon photoexcitation, and to generate design principles for the generation of light-induced quantum coherent states and their application in material-based quantum information technologies.

\noindent\textbf{Theory}

The time evolution of the population of electron-hole pairs $\alpha$, composing the many-body excitonic states, is computed using the density matrix $\rho$:

\begin{equation}\label{eq:dm_td}
\begin{split}
    \frac{d\rho_{\alpha_i,\alpha_j}}{dt}=\frac{1}{2}\sum_{\alpha'\alpha_i'\alpha_j'}[(\delta_{\alpha_i\alpha'}-\rho_{\alpha_i\alpha'})P^{eh}_{\alpha'\alpha_j,\alpha_i'\alpha_j'}\rho_{\alpha_i'\alpha_j'}\\
    -(\delta_{\alpha'\alpha_i'}-\rho_{\alpha'\alpha'_i})P^{eh*}_{\alpha'\alpha_i',\alpha_i\alpha_j'}\rho_{\alpha_j'\alpha_j}]+\mathrm{H.c.}.
\end{split}
\end{equation}
Here, $P^{eh}$ is the scattering superoperator describing the interactions between electron-hole pairs composing the many-body excitations, given by:
\begin{equation}\label{eq:P_tot}
P^{eh}=P^{ph-eh}+P^{Coul-eh}.
\end{equation}
The interaction of the electron-hole pairs with phonons is given in the first term, $P^{ph-eh}$. The second term, $P^{Coul-eh}$, includes Coulomb electron-electron and hole-hole scattering. 
We omit the indices from the subscripts in Eq.~\ref{eq:P_tot} for brevity. 
The Coulomb scattering is accounted by expanding a single-particle model-Hamiltonian scheme, previously derived by Rosati et al.~\cite{rosati2014derivation}, to include \textit{ab initio} many-body effects within the electron-hole basis set:

\begin{equation}\label{eq:P_col}
\begin{split}
{P^{coul-eh}_{eh_1eh_2eh_1'eh_2'}=\frac{4\pi\xi}{\hbar}\sum_{\bar{eh_1},\bar{eh_2},\bar{eh_1'},\bar{eh_2'}}\left(\delta_{\bar{eh_2}\bar{eh_1}}-\rho_{\bar{eh_2}\bar{eh_1}}\right)}
\\
{\times \hat{V}_{{eh_1}\bar{eh_1},{eh_1'}\bar{eh_1'}}\hat{V}^*_{{eh_2}\bar{eh_2},{eh_2'}\bar{eh_2'}}\rho_{\bar{eh_1'}\bar{eh_2'}}},
\end{split}
\end{equation}
here $eh_i$ refers to the $ith$ electron($e$)-hole($h$) pair, and $\xi$ is a matrix that enforces strict momentum conservation between initial and final states and a Gaussian-broadened energy conservation (see SI for further elaboration). The Coulomb pair-coupling terms are calculated via:

\begin{equation}\label{eq:V_sum}
\begin{split}
{\hat{V}_{eh\bar{eh}eh'\bar{eh'}}=\frac{1}{4}(\bar{V}_{eh\bar{eh}eh'\bar{eh'}} - \bar{V}_{\bar{eh}eheh'\bar{eh'}}}\\{ - \bar{V}_{eh\bar{eh}\bar{eh'}eh'} + \bar{V}_{\bar{eh}eh\bar{eh'}eh'})},
\end{split}
\end{equation}
where, $\bar{V}_{eh\bar{eh}eh'\bar{eh'}}$ includes electron-electron and hole-hole 
direct bare Coulomb interaction $V$:

\begin{equation}\label{eq:V_bar}
\begin{aligned}
\bar{V}_{eh\bar{eh}eh'\bar{eh'}} &= V_{e\bar{e}e'\bar{e'}}\delta_{hh'}\delta_{\bar{h}\bar{h'}}+ V_{h\bar{h}h'\bar{h'}}\delta_{ee'}\delta_{\bar{e}\bar{e'}}.
\end{aligned}
\end{equation}

The explicit inclusion of Coulomb interactions significantly enhances the computational complexity of the calculation, since the propagator superoperator becomes time-dependent through the current density matrix and has to be calculated at each time step during the propagation process. To allow longer time steps, we employ a fourth-order Runge-Kutta (RK4) method in the time propagation.

The electron-hole-pair - phonon scattering superoperator, which accounts for phonon scattering between electrons and holes, is given by:

\begin{equation}\label{eq:P_eh_ph}
\begin{split}
{P^{ph-eh}_{{eh}_1{eh}_2,{eh}_1'{eh}_2'}=\sum_{\pm,\nu}{(B^{\nu\pm}_{e_1e_1'h_1h_1}{B^{\nu\pm*}_{e_2e_2'h_2h_2}}\delta_{h_1h_1'}\delta_{h_2h_2'}}}
\\{+B^{\nu\pm}_{h_1h_1'e_1e_1}{B^{\nu\pm*}_{h_2h_2'e_2e_2}}\delta_{e_1e_1'}\delta_{e_2e_2'}}
\\{+B^{\nu\pm}_{e_1e_1'h_1h_1'}{B^{\nu\pm*}_{h_2h_2'e_2e_2'}}\delta_{h_1h_1'}\delta_{e_2e_2'}}
\\{+B^{\nu\pm}_{h_1h_1'e_1e_1'}{B^{\nu\pm*}_{e_2e_2'h_2h_2'}}\delta_{e_1e_1'}\delta_{h_2h_2'}}),
\end{split}
\end{equation}

with the phonon superoperators:

\begin{equation}\label{eq:B}
B^{\nu\pm}_{pp'nn'}=\sqrt{\frac{2\pi\left(f_{q\nu}+\frac{1}{2}\pm\frac{1}{2}\right)}{\hbar}}g^{\nu}_{pp'}D^{\nu\pm}_{pp'nn'}.
\end{equation}

$f_{q\nu}$ is the Bose-Einstein occupation of phonons with momentum $q$ and mode $\nu$ and $g$ the electron-phonon / hole-phonon coupling elements calculated using DFPT. $p$ and $p'$ stand for the interacting particles and $n$, $n'$ are non-interacting particles. $D$ is the energy conservation term, broadened by a Gaussian function:

\begin{equation}\label{eq:D}
D^{\nu\pm}_{pp'nn'}=\Delta\left(\epsilon_p-\epsilon_{p'}+\epsilon_n-\epsilon_{n'}\pm\hbar\omega_{q\nu}+E^c_{np}-E^c_{n'p'}\right).
\end{equation}

$\epsilon_i$ is the quasiparticle energy of state $i$, $\omega_{q\nu}$ is the phonon frequency calculated using DFPT and $E^c_{ij}$ the Coulomb coupling energy between quasiparticle states $i$ and $j$, obtained from elements of the BSE electron-hole interaction kernel. 
In the diagonal limit, the first two lines of Eq.~\ref{eq:P_eh_ph} are associated with the Fan-Migdal term and the last two lines with phonon-exchange.

To calculate excitonic lifetimes, we also include exciton recombination. For this, we follow previously developed methods~\cite{dolcini2013interplay, hod2016driven, zelovich2016driven, zelovich2017parameter} describing density matrices upon their relaxation back to equilibrium. We include state-dependent relaxation times calculated from \textit{ab initio}, which are time-independent and only non-zero for optically active electron-hole pairs, namely with exciton momentum $Q=0$.
In practice, we add another decay term to the propagation equation:
\begin{equation}
    \label{dm_to_eq}
    \frac{\partial \rho_{rec}}{\partial t}=-\Gamma \left(\rho-\rho_{eq}\right)=-\Gamma \rho,
\end{equation}
where $\Gamma$ is the \textit{ab initio}-calculated radiative decay time of each electron-hole pair, and is zero by definition for $Q\neq0$ pairs. The last equality assumes that the ground state is non-excitonic, hence $\rho_{eq}=0$, leading to a decrease in the trace of the density matrix over time.

To simulate an experimental setup, we define the initial density matrix to be a superposition of many-body exciton states. We do this by taking a linear combination of the different BSE solutions ($\left|\psi_n\right>$) with corresponding expansion coefficients ($C_n$), $\left|\Psi\right> =\sum_n {C_n\left|\psi_n\right>}$, and then expanding them to a density matrix form. The spanning coefficients are obtained by simulating a Gaussian pulse around an excitation energy with a chosen broadening, weighed by the oscillator strength of each excitonic state (see SI for further elaboration).

\noindent\textbf{Results}

We demonstrate the above-derived formalism for the case of monolayer MoSe$_2$. The propagation calculations are carried at 300K temperature and for the crystal structures relaxed using density functional theory (DFT) within the Quantum Epsresso~\cite{QE-2009, QE-2017, giannozzi2020quantum} code, then calculate quasiparticle and exciton states from GW-BSE using the BerkeleyGW package~\cite{deslippe2012berkeleygw}, as described in the SI~\cite{supp1}. As we demonstrate below, for this system, Coulomb interactions only induce small changes in the interaction picture. We thus first explore the change in exciton energy and spectral signature due to the phonon scattering channels and radiative recombination. Then, we study how the inclusion of Coulomb effects influence exciton coherence. Notably, for different systems with stronger Coulomb coupling, such as TMD heterostructures, the explicit inclusions of Coulomb interactions is expected to have more significant influence also on spectral properties. Importantly, the method derived here allows us this explicit examination of Coulomb interaction strength as a function of the system structure.

\begin{figure}
\includegraphics[width=1.0\linewidth]{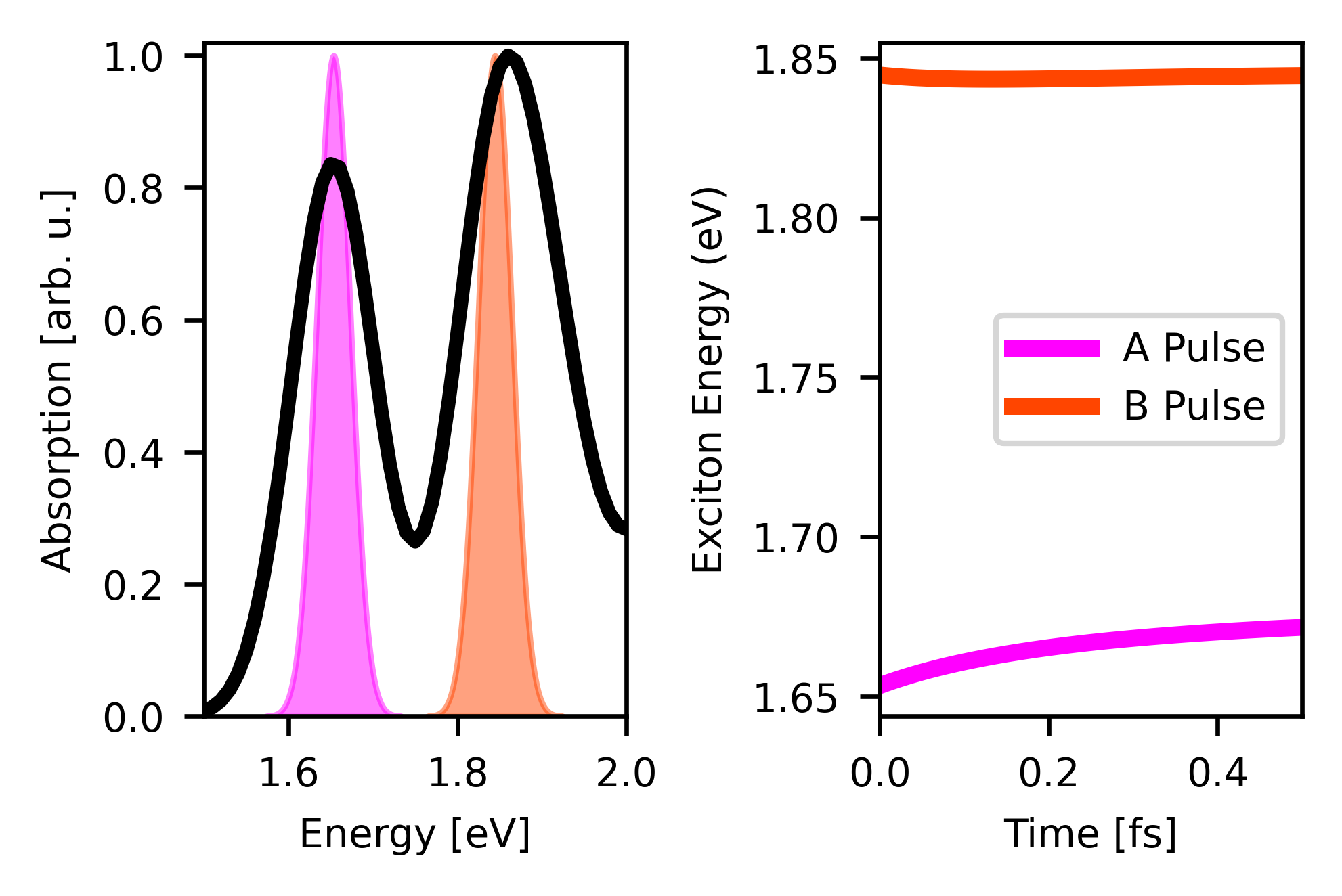}
\caption{(left) GW-BSE computed absorption spectrum (black), where each excitonic resonance is broadened by 50 meV, with two different initial Gaussian pulses for the dynamic calculations. The pink (orange) pulse is centered around around the lowest energy bright \textit{A} (\textit{B}) excitation.
(right) Exciton energy as a function of time since excitation, for the two initial pulses analyzed.}\label{fig:1}
\end{figure}

The left panel of Fig.~\ref{fig:1} presents the GW-BSE computed absorption spectrum. Two initial pulses are taken as starting points in the density matrix calculations of the dynamics, shown in pink (\textit{A} peak) and orange (\textit{B} peak). The change in the exciton energy of these two excitations due to phonon scattering and radiative recombination along the propagation time are shown in the right panel of Fig.~\ref{fig:1}, showing a gradual change in the $A$ peak energy, while the $B$ peak energy remains stable. This different behavior of the two pulses is dominated by both the initial conditions and the scattering terms included in the propagation. The most dominant elements of the energy change stem from the excitonic system interacting with the phonon bath, giving rise to a change in the energy of the system over time. The radiative decay processes, included in the propagation, are not dominant in the studied timescales, where the phonon interactions are four orders of magnitudes faster than the radiative decay rate.

\begin{figure*}
\includegraphics[width=0.82\linewidth]{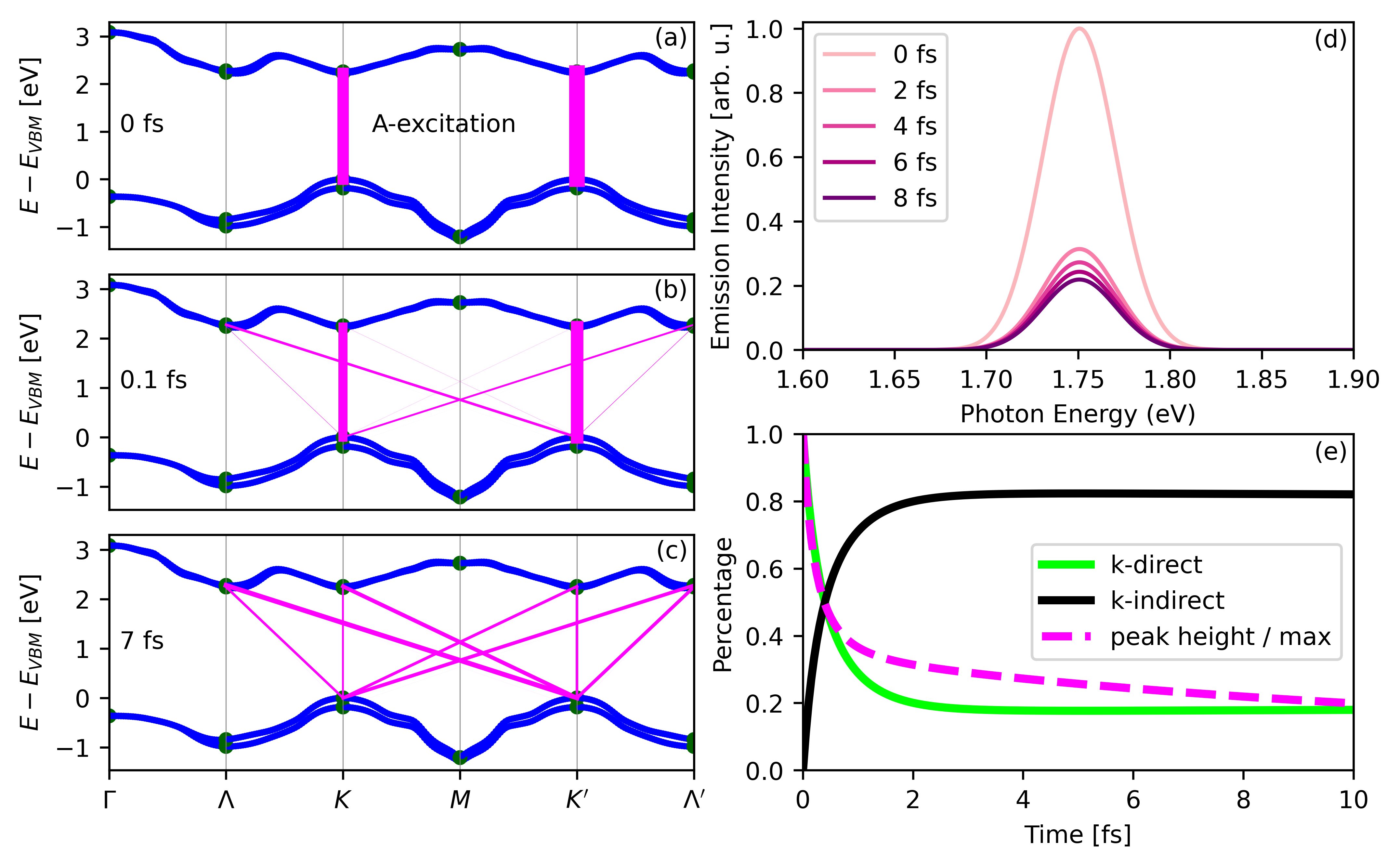}
\caption{(a) Electron/hole momentum points that participate in the propagation-grid, shown (in green) over the GW-computed quasiparticle bandstructure (blue). Magenta lines emphasize electron-hole pairs that build the initial excitonic state, centered around the \textit{A} peak. Line widths corresponds to the electron-hole pair relative weight within the many-body excitonic state. (b-c) Same as (a), for electron-hole pair weights after 0.1 (b) and after 7 (c) femtoseconds. (d) Computed time-dependent changes in the spectral height and width of the initial \textit{A} excitation, showing a reduction in the emission strength over time. (e) Time evolution of the emission peak height, relative to the maximum height which is at time zero (dashed magenta line), alongside the percentage of momentum direct (green) and momentum indirect (black) electron-hole transitions within the excited state.}\label{fig:2}
\end{figure*}

We further analyze the time-resolved changes in the relative population weights of different electron-hole pairs composing the excitonic states. Fig.~\ref{fig:2}(a)-(c) shows the propagation of an initial state centered around the \textit{A} peak where the magenta line widths, corresponding to the weights of the electron-hole pairs within the exciton state, vary over time. Ultrafast thermalization of the initial BSE excited state gives rise to the reduction of the spectral width, shown in Fig.~\ref{fig:2}(d), as optically-inactive, momentum indirect states populate over time, along with the reduction of the k-direct nature of the exciton, shown in Fig.~\ref{fig:2}(e). We note that using a finer grid in the propagation could lead to a less monotonically changing emission spectrum, although the general effect is expected to remain.

\begin{figure}
\includegraphics[width=1.0\linewidth]{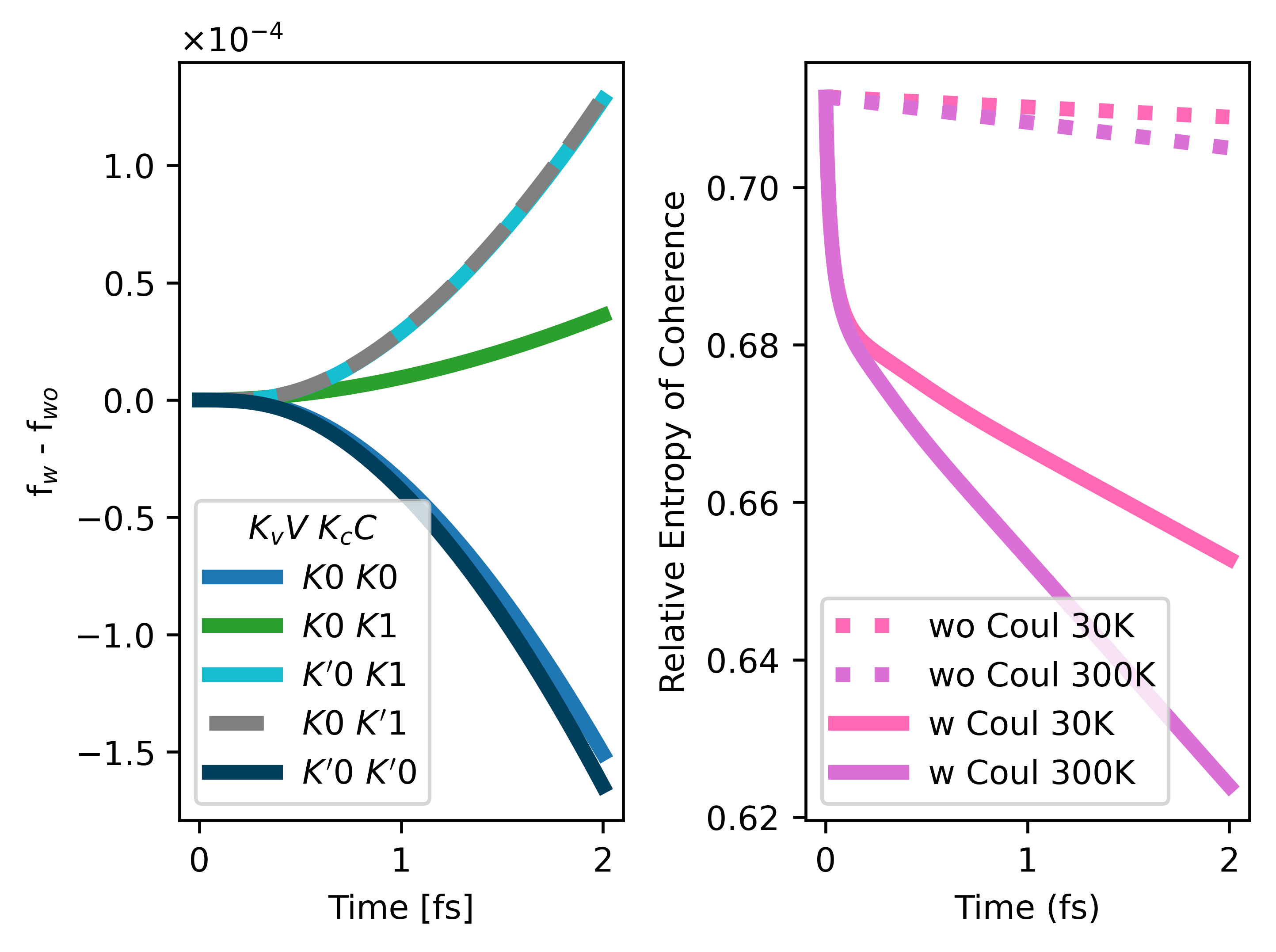}
\caption{(left) Population difference between electron-hole pair states within the propagating exciton, comparing simulations with ($f_\mathrm{W})$) and without ($f_{\mathrm{WO}})$) Coulomb scattering, as a function of time. Both simulations are performed at 300~K. (right) Relative entropy of coherence, defined in Eq.~\ref{eq:coh}, for an initially excited A-exciton state, shown over time at two temperatures (30~K and 300~K), with and without Coulomb scattering.}\label{fig:wwo}
\end{figure}

Finally, we examine the effect of the above-derived Coulomb scattering terms, Eq.~\ref{eq:P_col}, on the propagation dynamics, as shown in Fig.~\ref{fig:wwo}. The left panel displays the difference in state occupation between two simulations: one with explicit inclusion of Coulomb scattering and one without it (with occupation denoted as $f_w$ and $f_{wo}$, respectively). Due to the high computational cost of incorporating the time-dependent Coulomb scattering propagator, these simulations are performed on a reduced grid that includes only three high-symmetry points.
The differences in state occupation, represented by the diagonal elements of the density matrix, are minimal. It is essential to ensure that the simulation parameters preserve the physicality of the density matrix, such as maintaining unit trace and non-negative eigenvalues. Since including the radiative decay using the presented formalism does not conserve the density matrix trace, in this part we exclude it.

The right panel shows the relative entropy of coherence, $C_{\text{rel}}(\rho)$, quantifying the coherence of the density matrix over time~\cite{bu2017maximum, baumgratz2014quantifying}, computed via:
\begin{equation}\label{eq:coh}
C_{\text{rel}}(\rho) = S(\rho_{\text{diag}}) - S(\rho).
\end{equation}
Here, $S(\rho) = -\mathrm{Tr}(\rho \log \rho)$ is the von Neumann entropy of the state $\rho$, and $\rho_{\text{diag}}$ is the dephased state obtained by removing all off-diagonal elements of $\rho$ (see SI for further details).

Interestingly, despite the small differences in the diagonal elements upon introducing Coulomb scattering, the off-diagonal elements change significantly, leading to a pronounced reduction in state coherence. This behavior is observed at both low (30 K) and high (300 K) temperatures, highlighting the important interplay between temperature-dependent phonon scattering and temperature-independent Coulomb interactions—where the latter are implicitly modulated by phonon-induced changes in the quantum state, as also discussed from a different prescriptive in Ref.~\cite{perfetto2023real}.

\noindent\textbf{Discussion}

To summarize, in this work we presented an \textit{ab initio} approach to study exciton dynamics, demonstrated on monolayer MoSe$_2$, incorporating phonon-mediated scattering, Coulomb interactions, and radiative recombination within a Lindblad density matrix framework. Our methodology extends beyond conventional GW-BSE approaches by explicitly capturing ultrafast relaxation processes and quantum many-body interactions. Through time-dependent simulations, we examine the influence of different excitations on exciton relaxation pathways, demonstrating the critical role of phonon scattering in energy dissipation and the population of optically inactive states, as well as the effect of Coulomb interactions on exciton coherence dynamics. Our findings provide comprehensive understanding of exciton transport and recombination in the examined system. Our presented method is general and offers a predictive simulation approach for exploring exciton-driven processes in low-dimensional semiconductors and the design of next-generation optoelectronic and quantum materials.

\noindent{\textbf{Methods}}

First-principles calculations were performed to study monolayer MoSe$_2$, using DFT within the PBE approximation as implemented in Quantum Espresso~\cite{QE-2009,QE-2017,giannozzi2020quantum}, with norm-conserving pseudopotentials and spin-orbit coupling. Quasiparticle corrections and excitonic properties were computed using the GW-BSE approach with BerkeleyGW~\cite{deslippe2012berkeleygw}, including advanced sampling schemes~\cite{jordana2017nonuniform} to ensure convergence. Phonon spectra and electron-phonon coupling were obtained via density functional perturbation theory~\cite{giustino2017electron} using EPW~\cite{ponce2016epw}. Time-dependent density matrix propagation was carried out at finite temperature, including phonon and Coulomb scattering effects using home-written code. Full computational parameters and convergence details are provided in the Supplementary Materials.

\noindent{\textbf{Data Availability}}
DFT and GW-BSE data, as well as the dynamical and analysis data composing the results
presented in this study are available upon request from the corresponding authors.

\noindent{\textbf{Code Availability}}
DFT calculations are publicly-available within the QuantumEspresso code, see https://www.quantum-espresso.org/.
GW-BSE computations are publicly-available within the BerkeleyGW code, see https://berkeleygw.org/.
Density-matrix dynamics code will be published in a future version of BerkeleyGW.

\noindent{\textbf{Acknowledgments}} T. A. acknowledges support from the Azrieli Graduate Fellows Program. G. V.  acknowledges support from a Minerva Foundation Grant (No. 7135421). S. R.-A.  acknowledges support from an Israel Science Foundation Grant (No. 1208/19), and a European Research Council (ERC) Starting Grant (No. 101041159). M. D.B. acknowledges support by the Center for Computational Study of Excited-State Phenomena in Energy Materials (C2SEPEM), funded by the U.S. Department of Energy (DOE), Office of Science (SC), Basic Energy Sciences (BES), under contract No. DE-AC02-05CH11231. 
Computational resources were provided by the ChemFarm local cluster at the Weizmann Institute of Science.

\noindent{\textbf{Author Contributions}}
G. V., T. A. and S .R.A. developed the dynamics method. T. A. performed the computations. T. A. and S. R.A. analyzed the data. M. D.B. took part in the algorithmic and code developments. All authors contributed to the manuscript preparation.

\noindent{\textbf{Competing Interests}}
The authors declare no competing interests.

\noindent{\textbf{Correspondence}}
Correspondence and requests for materials should be addressed to Sivan Refaely-Abramson.

\bibliographystyle{unsrt}
\bibliography{references}

\begin{thebibliography}{10}

\bibitem{CohenLouie}
Marvin~L. Cohen and Steven~G. Louie.
\newblock {\em Fundamentals of Condensed Matter Physics}.
\newblock Cambridge University Press, 2016.

\bibitem{QuantumComputing}
Michael~A. Nielsen and Isaac~L. Chuang.
\newblock {\em Quantum Computation and Quantum Information}.
\newblock Cambridge University Press, 2010.

\bibitem{kennes2021moire}
Dante~M Kennes, Martin Claassen, Lede Xian, Antoine Georges, Andrew~J Millis, James Hone, Cory~R Dean, DN~Basov, Abhay~N Pasupathy, and Angel Rubio.
\newblock Moir{\'e} heterostructures as a condensed-matter quantum simulator.
\newblock {\em Nat. Phys.}, 17(2):155--163, 2021.

\bibitem{fogler2014high}
MM~Fogler, LV~Butov, and KS~Novoselov.
\newblock High-temperature superfluidity with indirect excitons in van der waals heterostructures.
\newblock {\em Nat. Commun.}, 5(1):4555, 2014.

\bibitem{katzer2023exciton}
Manuel Katzer, Malte Selig, Lukas Sigl, Mirco Troue, Johannes Figueiredo, Jonas Kiemle, Florian Sigger, Ursula Wurstbauer, Alexander~W. Holleitner, and Andreas Knorr.
\newblock Exciton-phonon scattering: Competition between the bosonic and fermionic nature of bound electron-hole pairs.
\newblock {\em Phys. Rev. B}, 108:L121102, Sep 2023.

\bibitem{qiu2013optical}
Diana~Y Qiu, Felipe~H Da~Jornada, and Steven~G Louie.
\newblock Optical spectrum of mos 2: many-body effects and diversity of exciton states.
\newblock {\em Phys. Rev. Lett.}, 111(21):216805, 2013.

\bibitem{berkelbach2018}
Timothy~C Berkelbach and David~R Reichman.
\newblock Optical and excitonic properties of atomically thin transition-metal dichalcogenides.
\newblock {\em Annu. Rev. Condens. Matter Phys.}, 9:379--396, 2018.

\bibitem{Ginsberg2020}
Naomi~S Ginsberg and William~A Tisdale.
\newblock Spatially resolved photogenerated exciton and charge transport in emerging semiconductors.
\newblock {\em Annu. Rev. Phys. Chem.}, 71:1--30, 2020.

\bibitem{Jin2018}
Chenhao Jin, Eric~Yue Ma, Ouri Karni, Emma~C. Regan, Feng Wang, and Tony~F. Heinz.
\newblock Ultrafast dynamics in van der waals heterostructures.
\newblock {\em Nat. Nanotechnol.}, 12:994--1003, 2018.

\bibitem{yuan2020twist}
Long Yuan, Biyuan Zheng, Jens Kunstmann, Thomas Brumme, Agnieszka~Beata Kuc, Chao Ma, Shibin Deng, Daria Blach, Anlian Pan, and Libai Huang.
\newblock Twist-angle-dependent interlayer exciton diffusion in ws2--wse2 heterobilayers.
\newblock {\em Nat. Mater.}, 19(6):617--623, 2020.

\bibitem{mitterreiter2021role}
Elmar Mitterreiter, Bruno Schuler, Ana Micevic, Daniel Hernang{\'o}mez-P{\'e}rez, Katja Barthelmi, Katherine~A Cochrane, Jonas Kiemle, Florian Sigger, Julian Klein, Edward Wong, Edward~S. Barnard, Kenji Watanabe, Takashi Taniguchi, Michael Lorke, Frank Jahnke, Johnathan~J. Finley, Adam~M. Schwartzberg, Diana~Y. Qiu, Sivan Refaely-Abramson, Alexander~W. Holleitner, Alexander Weber-Bargioni, and Christoph Kastl.
\newblock The role of chalcogen vacancies for atomic defect emission in mos2.
\newblock {\em Nat. Commun.}, 12(1):1--8, 2021.

\bibitem{hotger2023spin}
Alexander H{\"o}tger, Tomer Amit, Julian Klein, Katja Barthelmi, Thomas Pelini, Alex Delhomme, Sergio Rey, Marek Potemski, Cl{\'e}ment Faugeras, Galit Cohen, Daniel Hernangómez-Pérez, Takashi Taniguchi, Kenji Watanabe, Christoph Kastl, Jonathan~J Finley, Sivan Refaely-Abramson, Alexander~W Holleitner, and Andreas~V Stier.
\newblock Spin-defect characteristics of single sulfur vacancies in monolayer mos2.
\newblock {\em npj 2D Mater. Appl.}, 7(1):30, 2023.

\bibitem{troue2023extended}
Mirco Troue, Johannes Figueiredo, Lukas Sigl, Christos Paspalides, Manuel Katzer, Takashi Taniguchi, Kenji Watanabe, Malte Selig, Andreas Knorr, Ursula Wurstbauer, and Alexander~W. Holleitner.
\newblock Extended spatial coherence of interlayer excitons in ${\mathrm{mose}}_{2}/{\mathrm{wse}}_{2}$ heterobilayers.
\newblock {\em Phys. Rev. Lett.}, 131:036902, Jul 2023.

\bibitem{refaely2018defect}
Sivan Refaely-Abramson, Diana~Y Qiu, Steven~G Louie, and Jeffrey~B Neaton.
\newblock Defect-induced modification of low-lying excitons and valley selectivity in monolayer transition metal dichalcogenides.
\newblock {\em Phys. Rev. Lett.}, 121(16):167402, 2018.

\bibitem{amit2022tunable}
Tomer Amit, Daniel Hernang{\'o}mez-P{\'e}rez, Galit Cohen, Diana~Y Qiu, and Sivan Refaely-Abramson.
\newblock Tunable magneto-optical properties in mos 2 via defect-induced exciton transitions.
\newblock {\em Phys. Rev. B}, 106(16):L161407, 2022.

\bibitem{kundu2023exciton}
Sudipta Kundu, Tomer Amit, HR~Krishnamurthy, Manish Jain, and Sivan Refaely-Abramson.
\newblock Exciton fine structure in twisted transition metal dichalcogenide heterostructures.
\newblock {\em Npj Comput. Mater.}, 9(1):186, 2023.

\bibitem{Qiu2021}
Diana~Y Qiu, Galit Cohen, Dana Novichkova, and Sivan Refaely-Abramson.
\newblock Signatures of dimensionality and symmetry in exciton band structure: Consequences for exciton dynamics and transport.
\newblock {\em Nano Lett.}, 21(18):7644--7650, 2021.

\bibitem{Hybertsen1986}
M.~S. Hybertsen and S.~G. Louie.
\newblock Electron correlation in semiconductors and insulators: Band gaps and quasiparticle energies.
\newblock {\em Phys. Rev. B}, 34:5390, 1986.

\bibitem{Rohlfing1998}
M.~Rohlfing and S.~G. Louie.
\newblock Electron-hole excitations in semiconductors and insulators.
\newblock {\em Phys. Rev. Lett.}, 81:2312, 1998.

\bibitem{Qiu2015}
D.~Y. Qiu, T.~Cao, and S.~G. Louie.
\newblock Nonanalyticity, valley quantum phases, and lightlike exciton dispersion in monolayer transition metal dichalcogenides: Theory and first-principles calculations.
\newblock {\em Phys. Rev. Lett.}, 115:176801, 2015.

\bibitem{Cudazzo2015}
Pierluigi Cudazzo, Francesco Sottile, Angel Rubio, and Matteo Gatti.
\newblock Exciton dispersion in molecular solids.
\newblock {\em J. Phys.: Condens. Matter}, 27:113204, 2015.

\bibitem{louie2021discovering}
Steven~G Louie, Yang-Hao Chan, Felipe~H da~Jornada, Zhenglu Li, and Diana~Y Qiu.
\newblock Discovering and understanding materials through computation.
\newblock {\em Nat. Mater.}, 20(6):728--735, 2021.

\bibitem{Attaccalite2011}
C.~Attaccalite, M.~Gr\"{u}ning, and A.~Marini.
\newblock Real-time approach to the optical properties of solids and nanostructures: Time-dependent bethe-salpeter equation.
\newblock {\em Phys. Rev. B}, 84:245110, 2011.

\bibitem{perfetto2022real}
Enrico Perfetto, Yaroslav Pavlyukh, and Gianluca Stefanucci.
\newblock Real-time g w: Toward an ab initio description of the ultrafast carrier and exciton dynamics in two-dimensional materials.
\newblock {\em Phys. Rev. Lett.}, 128(1):016801, 2022.

\bibitem{perfetto2023real}
Enrico Perfetto and Gianluca Stefanucci.
\newblock Real-time gw-ehrenfest-fan-migdal method for nonequilibrium 2d materials.
\newblock {\em Nano Lett.}, 23(15):7029--7036, 2023.

\bibitem{perfetto2015nonequilibrium}
E~Perfetto, D~Sangalli, A~Marini, and G~Stefanucci.
\newblock Nonequilibrium bethe-salpeter equation for transient photoabsorption spectroscopy.
\newblock {\em Phys. Rev. B}, 92(20):205304, 2015.

\bibitem{sangalli2018ab}
Davide Sangalli, Enrico Perfetto, Gianluca Stefanucci, and Andrea Marini.
\newblock An ab-initio approach to describe coherent and non-coherent exciton dynamics.
\newblock {\em Euro. Phys. J. B}, 91(8):1--12, 2018.

\bibitem{chan2021giant}
Yang-Hao Chan, Diana~Y Qiu, Felipe~H da~Jornada, and Steven~G Louie.
\newblock Giant exciton-enhanced shift currents and direct current conduction with subbandgap photo excitations produced by many-electron interactions.
\newblock {\em Proc. Nat. Acad. Sci.}, 118(25):e1906938118, 2021.

\bibitem{sangalli2021excitons}
D~Sangalli.
\newblock Excitons and carriers in transient absorption and time-resolved arpes spectroscopy: An ab initio approach.
\newblock {\em Phys. Rev. Mater.}, 5(8):083803, 2021.

\bibitem{antonius2022theory}
Gabriel Antonius and Steven~G Louie.
\newblock Theory of exciton-phonon coupling.
\newblock {\em Phys. Rev. B}, 105(8):085111, 2022.

\bibitem{chen2022first}
Hsiao-Yi Chen, Davide Sangalli, and Marco Bernardi.
\newblock First-principles ultrafast exciton dynamics and time-domain spectroscopies: Dark-exciton mediated valley depolarization in monolayer wse2.
\newblock {\em Phys. Rev. Res.}, 4(4):043203, 2022.

\bibitem{chan2023exciton}
Yang-hao Chan, Jonah~B Haber, Mit~H Naik, Jeffrey~B Neaton, Diana~Y Qiu, Felipe~H da~Jornada, and Steven~G Louie.
\newblock Exciton lifetime and optical line width profile via exciton--phonon interactions: Theory and first-principles calculations for monolayer mos2.
\newblock {\em Nano Lett.}, 23(9):3971--3977, 2023.

\bibitem{cohen2024phonon}
Galit Cohen, Jonah~B Haber, Jeffrey~B Neaton, Diana~Y Qiu, and Sivan Refaely-Abramson.
\newblock Phonon-driven femtosecond dynamics of excitons in crystalline pentacene from first principles.
\newblock {\em Phys. Rev. Lett.}, 132(12):126902, 2024.

\bibitem{giustino2017electron}
Feliciano Giustino.
\newblock Electron-phonon interactions from first principles.
\newblock {\em Rev. Mod. Phys.}, 89(1):015003, 2017.

\bibitem{amit2023ultrafast}
Tomer Amit and Sivan Refaely-Abramson.
\newblock Ultrafast exciton decomposition in transition metal dichalcogenide heterostructures.
\newblock {\em Phys. Rev. B}, 108(22):L220305, 2023.

\bibitem{rosati2014derivation}
Roberto Rosati, Rita~Claudia Iotti, Fabrizio Dolcini, and Fausto Rossi.
\newblock Derivation of nonlinear single-particle equations via many-body lindblad superoperators: A density-matrix approach.
\newblock {\em Phys. Rev. B}, 90(12):125140, 2014.

\bibitem{dolcini2013interplay}
Fabrizio Dolcini, Rita~Claudia Iotti, and Fausto Rossi.
\newblock Interplay between energy dissipation and reservoir-induced thermalization in nonequilibrium quantum nanodevices.
\newblock {\em Phys. Rev. B}, 88(11):115421, 2013.

\bibitem{hod2016driven}
Oded Hod, C{\'e}sar~A Rodr{\'\i}guez-Rosario, Tamar Zelovich, and Thomas Frauenheim.
\newblock Driven liouville von neumann equation in lindblad form.
\newblock {\em J. Phys. Chem. A}, 120(19):3278--3285, 2016.

\bibitem{zelovich2016driven}
Tamar Zelovich, Leeor Kronik, and Oded Hod.
\newblock Driven liouville von neumann approach for time-dependent electronic transport calculations in a nonorthogonal basis-set representation.
\newblock {\em J. Phys. Chem. C}, 120(28):15052--15062, 2016.

\bibitem{zelovich2017parameter}
Tamar Zelovich, Thorsten Hansen, Zhen-Fei Liu, Jeffrey~B Neaton, Leeor Kronik, and Oded Hod.
\newblock Parameter-free driven liouville-von neumann approach for time-dependent electronic transport simulations in open quantum systems.
\newblock {\em J. Chem. Phys.}, 146(9), 2017.

\bibitem{QE-2009}
Paolo Giannozzi, Stefano Baroni, Nicola Bonini, Matteo Calandra, Roberto Car, Carlo Cavazzoni, Davide Ceresoli, Guido~L Chiarotti, Matteo Cococcioni, Ismaila Dabo, Andrea {Dal Corso}, Stefano de~Gironcoli, Stefano Fabris, Guido Fratesi, Ralph Gebauer, Uwe Gerstmann, Christos Gougoussis, Anton Kokalj, Michele Lazzeri, Layla Martin-Samos, Nicola Marzari, Francesco Mauri, Riccardo Mazzarello, Stefano Paolini, Alfredo Pasquarello, Lorenzo Paulatto, Carlo Sbraccia, Sandro Scandolo, Gabriele Sclauzero, Ari~P Seitsonen, Alexander Smogunov, Paolo Umari, and Renata~M Wentzcovitch.
\newblock Quantum espresso: a modular and open-source software project for quantum simulations of materials.
\newblock {\em J. Phys.: Condens. Matter}, 21(39):395502 (19pp), 2009.

\bibitem{QE-2017}
P~Giannozzi, O~Andreussi, T~Brumme, O~Bunau, M~Buongiorno Nardelli, M~Calandra, R~Car, C~Cavazzoni, D~Ceresoli, M~Cococcioni, N~Colonna, I~Carnimeo, A~Dal Corso, S~de~Gironcoli, P~Delugas, R~A~DiStasio Jr, A~Ferretti, A~Floris, G~Fratesi, G~Fugallo, R~Gebauer, U~Gerstmann, F~Giustino, T~Gorni, J~Jia, M~Kawamura, H-Y Ko, A~Kokalj, E~Küçükbenli, M~Lazzeri, M~Marsili, N~Marzari, F~Mauri, N~L Nguyen, H-V Nguyen, A~Otero de-la Roza, L~Paulatto, S~Poncé, D~Rocca, R~Sabatini, B~Santra, M~Schlipf, A~P Seitsonen, A~Smogunov, I~Timrov, T~Thonhauser, P~Umari, N~Vast, X~Wu, and S~Baroni.
\newblock Advanced capabilities for materials modelling with quantum espresso.
\newblock {\em J. Phys.: Condens. Matter}, 29(46):465901, 2017.

\bibitem{giannozzi2020quantum}
Paolo Giannozzi, Oscar Baseggio, Pietro Bonf{\`a}, Davide Brunato, Roberto Car, Ivan Carnimeo, Carlo Cavazzoni, Stefano De~Gironcoli, Pietro Delugas, Fabrizio Ferrari~Ruffino, et~al.
\newblock Quantum espresso toward the exascale.
\newblock {\em J. Chem. Phys.}, 152(15), 2020.

\bibitem{deslippe2012berkeleygw}
Jack Deslippe, Georgy Samsonidze, David~A Strubbe, Manish Jain, Marvin~L Cohen, and Steven~G Louie.
\newblock Berkeleygw: A massively parallel computer package for the calculation of the quasiparticle and optical properties of materials and nanostructures.
\newblock {\em Comput. Phys. Commun.}, 183(6):1269--1289, 2012.

\bibitem{supp1}
See Supplemental Material at [URL will be inserted by publisher] for more elaborate details about the computational method and computational details.

\bibitem{bu2017maximum}
Kaifeng Bu, Uttam Singh, Shao-Ming Fei, Arun~Kumar Pati, and Junde Wu.
\newblock Maximum relative entropy of coherence: an operational coherence measure.
\newblock {\em Phys. Rev. Lett.}, 119(15):150405, 2017.

\bibitem{baumgratz2014quantifying}
Tillmann Baumgratz, Marcus Cramer, and Martin~B Plenio.
\newblock Quantifying coherence.
\newblock {\em Phys. Rev. Lett.}, 113(14):140401, 2014.

\bibitem{jordana2017nonuniform}
Felipe~H. da~Jornada, Diana~Y. Qiu, and Steven~G. Louie.
\newblock {Nonuniform sampling schemes of the Brillouin zone for many-electron perturbation-theory calculations in reduced dimensionality}.
\newblock {\em Phys. Rev. B}, 95(3):035109, 2017.

\bibitem{ponce2016epw}
Samuel Ponc{\'e}, Elena~R Margine, Carla Verdi, and Feliciano Giustino.
\newblock Epw: Electron--phonon coupling, transport and superconducting properties using maximally localized wannier functions.
\newblock {\em Comput. Phys. Commun.}, 209:116--133, 2016.

\end{thebibliography}

\end{document}